\documentclass[twocolumn,aps,nofootinbib]{revtex4}
\usepackage{graphicx, epsfig, array}

\newcommand{\be}{\begin{equation}}
\newcommand{\ee}{\end{equation}}
\newcommand{\bea}{\begin{eqnarray}}
\newcommand{\eea}{\end{eqnarray}}

\newcommand{\ApJSS}{{Astrophys. J. Suppl.\ Ser.\,}}

\newcommand{\etal}{{\it et al.}}
\def\ibid#1#2#3{{\it ibid. }{\bf #1},~#3~(#2)}

\def\fun#1#2{\lower3.6pt\vbox{\baselineskip0pt\lineskip.9pt
        \ialign{$\mathsurround=0pt#1\hfill##\hfil$\crcr#2\crcr\sim\crcr}}}

\newcommand{\vis}{\text{vis}}
\newcommand\lsim{\mathrel{\rlap{\lower4pt\hbox{\hskip1pt$\sim$}}
    \raise1pt\hbox{$<$}}}
\newcommand\gsim{\mathrel{\rlap{\lower4pt\hbox{\hskip1pt$\sim$}}
    \raise1pt\hbox{$>$}}}

\def\dslash{\not{\hbox{\kern-2pt $\partial$}}}
\def\Dslash{\not{\hbox{\kern-4pt $D$}}}
\def\Oslash{\not{\hbox{\kern-4pt $O$}}}
\def\Qslash{\not{\hbox{\kern-4pt $Q$}}}
\def\pslash{\not{\hbox{\kern-2.3pt $p$}}}
\def\kslash{\not{\hbox{\kern-2.3pt $k$}}}
\def\qslash{\not{\hbox{\kern-2.3pt $q$}}}
 \newtoks\slashfraction
 \slashfraction={.13}
 \def\slash#1{\setbox0\hbox{$ #1 $}
 \setbox0\hbox to \the\slashfraction\wd0{\hss \box0}/\box0 }
\def\ee{\end{equation}}
\def\be{\begin{equation}}

\begin{document}
\setlength{\unitlength}{1mm}
\title{Indication for Primordial Anisotropies in the Neutrino Background from
WMAP and SDSS}
\author{Roberto Trotta$^{1,2}$,
Alessandro Melchiorri$^3$}
\address{
 $^1$ Oxford University, Astrophysics,  Denys Wilkinson Building,
Keble Road, Oxford, OX1 3RH, United Kingdom \\
 $^2$D\'epartement de Physique
Th\'eorique, Universit\'e de Gen\`eve,
24 quai Ernest Ansermet, CH-1211 Gen\`eve 4, Switzerland\\
 $^3$ Physics Department and sezione INFN, University of Rome ``La
Sapienza'',
Ple Aldo Moro 2, 00185 Rome, Italy\\
}
\date{\today}%
\begin{abstract}
We demonstrate that combining Cosmic Microwave Background
anisotropy measurements from the 1st year WMAP observations with
clustering data from the SLOAN galaxy redshift survey yields an
indication for primordial anisotropies in the cosmological
Neutrino Background.
\end{abstract}
\bigskip
\maketitle {\bf Introduction} Recent cosmological data coming from
measurements of the Cosmic Microwave Background (CMB) anisotropies
(see e.g. \cite{Bennett03}), on galaxy clustering (see e.g.
\cite{Tg04}) and, more recently, on Lyman-alpha Forest clouds (see
e.g. \cite{Se04}) are in spectacular agreement with the
expectations of the so-called standard model of structure
formation, based on primordial, purely adiabatic inflationary
perturbations and a cosmological constant (see e.g.
\cite{Se04},\cite{spergel}).

Cosmology is therefore becoming more and more a powerful
laboratory where physics not easily accessible on Earth can be
tested and verified. An excellent example of this comes from the
new cosmological constraints on neutrino physics. According to the
standard model, a cosmic Neutrino Background (NB), similar to the
CMB, should exist in our universe with a density of approximately
$n_{\nu}\sim 57 / \text{cm}^3$ per flavor. Neutrinos of mass $\ll
10^{-3} \text{eV}$ would still be relativistic today with a
Fermi-Dirac spectrum at a temperature $T \sim 2 \text{K}$, while
more massive neutrinos would be non relativistic and clustered
around galaxies with a typical velocity of $v_\nu \sim 200
\text{km/s}$. Therefore, despite the high density of cosmological
neutrinos a direct detection is virtually impossible due to their
extremely low energy and cross sections (see e.g. \cite{hagmann}).

However, cosmological neutrinos have a profound impact on
cosmology since they change the expansion history of the universe
and affect the growth of perturbations (see
\cite{Bashinsky:2003tk} for a detailed account). As a consequence,
recent cosmological data do provide strong -- albeit indirect --
evidence for the presence of a NB (see e.g. \cite{bowen}) and have
been used to put upper limits on absolute neutrino masses
competitive with those from laboratory experiments (see e.g.
\cite{Se04}, \cite{fogli}).
\begin{figure}[t]
\begin{center}
\includegraphics[angle=-90,width=1.05\linewidth]{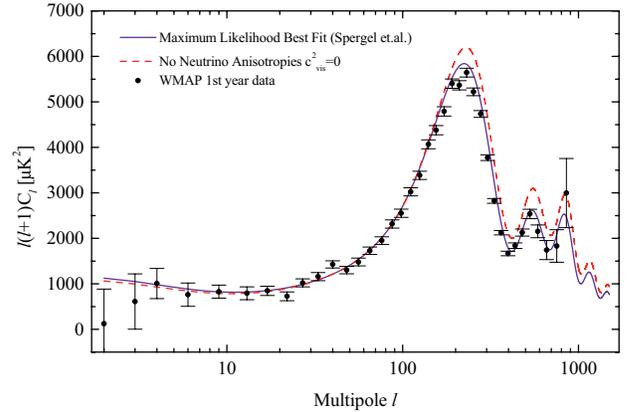}
\caption{The effect of NB anisotropies on the CMB temperature
angular power spectrum. The standard model with parameters which
provide the maximum likelihood best fit to WMAP is plotted (see
\cite{spergel}) against the same model but with no NB
anisotropies. The 1st year WMAP data is also plotted for
comparison.} \label{figcivs}
\end{center}
\end{figure}

In this {\it letter} we show that current cosmological data now
provide {\it for the first time}  an interesting indication for
primordial anisotropies in the NB. Although inflationary
anisotropies in the NB at the level of $\sim 10^{-5}$ are expected
in the standard scenario, a direct detection is clearly
impossible. However, anisotropies in the NB background affect the
CMB anisotropy angular power spectrum at level of $\sim 20 \%$
through the gravitational feedback of their free streaming damping
and anisotropic stress contributions \cite{Huetal95} and an
indirect detection is indeed possible.

{\bf Data analysis} A way to parameterize the anisotropies in the
NB has been introduced in \cite{Hu98} with the ``viscosity
parameter'' $c_{\vis}^2$, which controls the relationship between
velocity/metric shear and anisotropic stress in the NB. A value of
$c_{\vis}^2=1/3$ is what one expects in the standard scenario,
where anisotropies are present in the NB and approximate the
radiative viscosity of real neutrinos (see Fig.~1). The case
$c_{\vis}^2=0$, on the contrary, cuts the Boltzmann hierarchy of
NB perturbations at the quadrupole, forcing a perfect fluid
solution with no NB anisotropies but only density and velocity
(pressure) perturbations. Furthermore, the generation of
polarization is also modified, due to changes induced in the
polarization source term. Cosmological data which would
discriminate $c_{\vis}^2=0$ in favor of $c_{\vis}^2=1/3$ would
therefore provide a strong indication for the existence of NB
anisotropies, as argued in \cite{Huetal98}.

We show that this is indeed the case by computing the constraints
on $c_{\vis}^2$ obtained by combining several cosmological data
sets. The method we adopt is based on the publicly available
Markov Chain Monte Carlo package
\texttt{cosmomc}\footnote{Available from
\texttt{http://cosmologist.info}.} \cite{Lewis:2002ah}, which has
been modified to allow for values of $c_{\vis}^2 \neq 1/3$. We
sample the following 7 dimensional set of cosmological parameters,
adopting flat priors on them: $c_{\vis}^2$, the physical baryon
and CDM densities, $\omega_b=\Omega_bh^2$ and
$\omega_c=\Omega_ch^2$, the ratio of the sound horizon to the
angular diameter distance at decoupling, $\Theta_s$, the scalar
spectral index and the overall normalization of the spectrum,
$n_s$ and $A_s$, and, finally, the optical depth to reionization,
$\tau_r$. Furthermore, we consider purely adiabatic initial
conditions, we impose flatness and we do not include gravitational
waves. We first restrict our analysis to the case of 3 massless
neutrino families, thereby fixing the background neutrino density,
but we proceed to relax this assumption at the end. Introducing a
neutrino mass in agreement with current neutrino oscillation data
does not change our results in a significant way.

We include the first-year data \cite{wmap1} (temperature and
polarization) with the routine for computing the likelihood
supplied by the WMAP team \cite{verde}, as well as the CBI
\cite{cbi}, VSA \cite{vsa} and ACBAR \cite{acbar1,acbarpage}
measurements of the CMB. The MC convergence diagnostics is done
along the lines described in \cite{Lazarides:2004we}. In addition
to the CMB data, we also consider the constraints on the
real-space power spectrum of galaxies from the SLOAN galaxy
redshift survey (SDSS) \cite{sloan}. We restrict the analysis to a
range of scales over which the fluctuations are assumed to be in
the linear regime ($k < 0.2 h^{-1}\rm Mpc$). When combining the
matter power spectrum with CMB data, we marginalize over a bias
$b$ considered as an additional nuisance parameter. Furthermore,
we make use of the HST measurement of the Hubble parameter $H_0 =
100h \quad \text{km s}^{-1} \text{Mpc}^{-1}$ \cite{freedman} by
multiplying the likelihood by a Gaussian likelihood function
centered around $h=0.72$ and with a standard deviation $\sigma =
0.08$.
\begin{figure}[t]
\begin{center}
\includegraphics[width=0.8\linewidth]{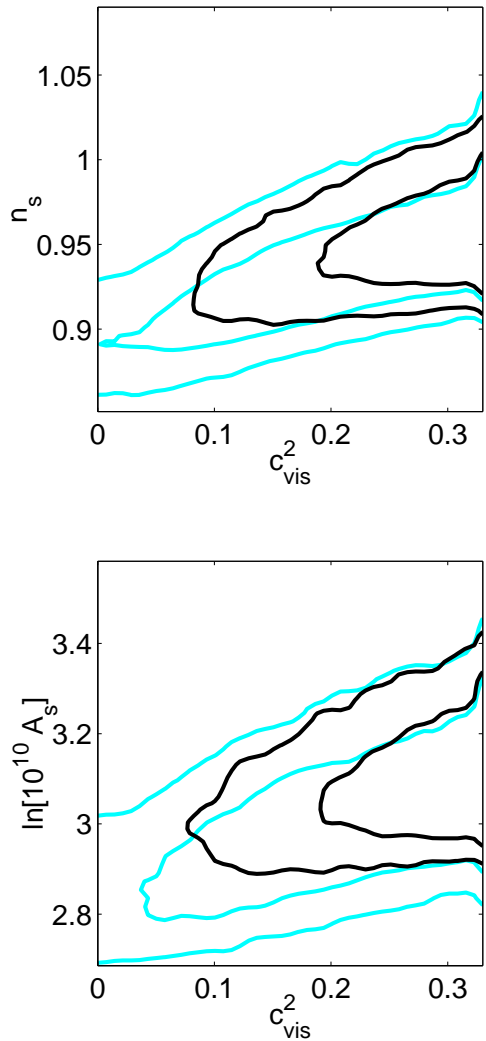}
\caption{Joint 2-dimensional posterior probability contour plots
in the $c_{\vis}^2-n_s$ (top) and $c_{\vis}^2-A_s$ (bottom)
planes, showing the $68\%$ and $95\%$ contours from the WMAP+other
CMB data alone (cyan/light gray) and adding SLOAN matter power
spectrum information (black). Degeneracies between these
parameters are evident and no lower limit on $c_{\vis}^2$ can be
placed with CMB data alone.} \label{figcivs}
\end{center}
\end{figure}

In Fig.~2 we plot the constraints obtained from our analysis in
the $c_{\vis}^2-n_s$ and $c_{\vis}^2-A_s$ planes. Including CMB
and matter power spectrum data, the best fit for the standard case
with $c_{\vis}^2=1/3$ has a $\chi^2 = 1482.9$, while for the
$c_{\vis}^2=0$ case we obtain $\chi^2=1490.2$, clearly favoring
the standard case ($\Delta \chi^2 = 7.3$). Furthermore, the best
fit model with $c_{\vis}^2 = 0$ has a rather tilted spectrum ($n_s
= 0.90$), while the combined effect of a red index and a slightly
smaller amplitude of the fluctuations reduces the optical depth to
$\tau_r = 0.07$. As already noticed in \cite{Huetal98}, it is in
fact possible to identify a correlation between the amplitude
$A_s$ and spectral index $n_s$ of the primordial fluctuations and
the value of $c_{\vis}^2$. This arises because the absence of NB
anisotropies boosts the amplitude of the CMB acoustic peaks, see
Fig.~1. This can be compensated, for example, with a lower value
for the spectral index $n_s$. The degeneracy is nearly exact for
CMB anisotropies alone and we found that no strong constraint can
be placed from the CMB data only: in that case, the best fit value
for the standard case is $\chi^2=1452.5$, while the $c_{\vis}^2=0$
case has $\chi^2=1454.5$. However, the latter case has as best fit
values $h=0.79$ and $n_s=0.89$, which are somewhat at odds with
the HST measurement and with the combined WMAP+SLOAN analysis,
which indicates a spectral index close to scale invariance. The
value of $c_{\vis}^2$ has very little impact on the matter power
spectrum, but inclusion of the SLOAN data does reduce the allowed
range of the other cosmological parameters, and especially of the
tilt, thereby allowing to place more stringent bounds on
$c_{\vis}^2$, as illustrated in Fig.~3. The 1D marginalized
posterior distribution for $c_{\vis}^2$ yields a lower limit
$c_{\vis}^2 > 0.12$ ($2\sigma$ c.l., 1 tail). The value
$c_{\vis}^2 = 0$ is found to lie about $2.4\sigma$ away from the
standard case, $c_{\vis}^2 = 1/3$. Therefore we can conclude that
the case where the NB does not have anisotropies above the first
moment is quite clearly disfavored.
\begin{figure}[t]
\begin{center}
\includegraphics[width=\linewidth]{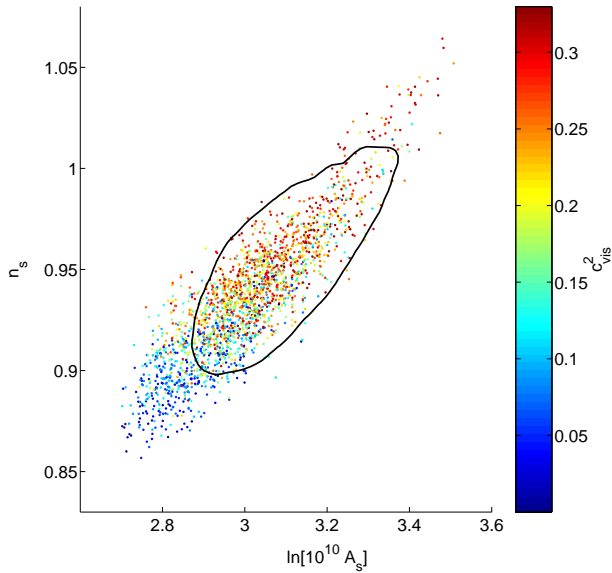}
\caption{Illustration of the role of galaxy clustering data in
constraining $c_{\vis}^2$. With CMB data only (scatter plot),
lower values of $c_{\vis}^2$ can be compensated by a redder
spectral index and a lower amplitude of the fluctuations.
Inclusion of the SLOAN data disfavors models with low $n_s$,
thereby cutting away most samples with low $c_{\vis}^2$. The solid
black contour is the joint $2\sigma$ posterior from CMB and
SLOAN.} \label{figcivs3D}
\end{center}
\end{figure}

\begin{figure}[t]
\begin{center}
\includegraphics[width=\linewidth]{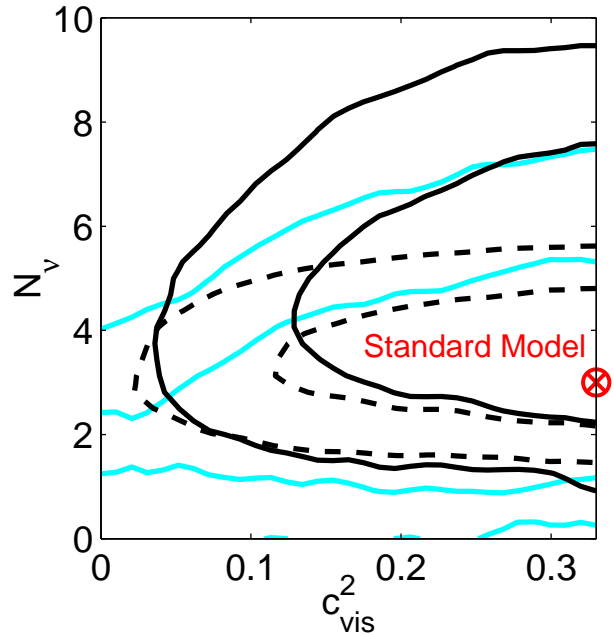}
\caption{Relaxing the assumption on the background neutrino
density by including the number of neutrino families $N_\nu$ as a
free parameter: joint 2-dimensional posterior probability contour
plots in the $c_{\vis}^2-N_\nu$ plane, enclosing $68\%$ and $95\%$
probability from the WMAP+other CMB data alone (cyan/light gray),
adding SLOAN matter power spectrum information (black, solid) and
further imposing a conservative BBN constraint on $N_\nu$ (black,
dashed). The indication that $c_{\vis}^2
> 0$ is largely independent on prior knowledge of $N_\nu$, since
the joint WMAP+SDSS data can constrain $c_{\vis}^2$ and $N_\nu$
simultaneously. The prediction of the Standard Model is shown by
the cross.} \label{figcivsNnu}
\end{center}
\end{figure}

Finally, it is interesting to relax the assumption on the number
of massless neutrino species, $N_\nu$, in order to test the
prediction of the Standard Model that $(c_{\vis}^2, N_\nu) = (1/3,
3)$. Departures from this point would constitute a strong
indication for the presence of new physics. By treating $N_\nu$ as
a free parameter in the range $0 \leq N_\nu \leq 10$ we obtain the
constraints depicted in Fig.~\ref{figcivsNnu}. Joint contours in
the $c_{\vis}^2 - N_\nu$ plane enclose the predicted value $(1/3,
3)$, and we do not find a significant degeneracy between the two
parameters, since their impact on the CMB is almost orthogonal. As
a consequence, the marginalized bound on $c_{\vis}^2$ is only
slightly weakened, and we obtain that $c_{\vis}^2=0$ is disfavored
at just above the $2\sigma$ level (CMB and SLOAN data). Further
imposing a conservative Big Bang Nucleosynthesis (BBN) constraint
on $N_\nu$ in the form of a Gaussian likelihood centered around
$N_\nu = 3$ and with spread $\Delta N_\nu = 1$ \cite{bbn} does not
change the result for $c_{\vis}^2$ in a significative way (see
Fig.~\ref{figcivsNnu}).

From a Bayesian model selection point of view, we can compare the
evidence in favor of $c_{\vis}^2=1/3$ as opposed to $c_{\vis}^2 <
1/3$ using the Savage-Dickey density ratio and taking a flat prior
in the range $0\leq c_{\vis}^2\leq 1/3$ (see \cite{robbays} for an
explanation of the method and precise definitions). We obtain that
WMAP+SDSS data favor $c_{\vis}^2 = 1/3$ with odds slightly larger
than $2:1$, irrespective of the assumptions on the neutrino
background density, constituting positive (if only weak) evidence
in favor of the Standard Model value.

We also considered models with $c_{\vis}^2>1/3$, but in this case
we found that the current data does not provide any relevant
constraint up to $c_{\vis}^2=1$.

{\bf Conclusions} In this {\em letter} we show that current
cosmological data are providing for the first time an interesting
indication for primordial anisotropies in the cosmological
neutrino background. Our result shows indication for the existence
of a NB presenting anisotropies as predicted by the standard
scenario, in which higher order multipoles in the neutrino
distribution function are generated by free streaming. The
significance is still small (slightly larger than $2 \sigma$,
quite independently on the assumptions on the radiation content)
and possibly plagued by systematics.

We analyzed a limited set of models and our conclusions are valid
only in the theoretical framework we consider. Enlarging the
cosmological model by increasing the number of parameters could
lower the significance of our result. For instance, a running of
the spectral index would make the degeneracy between $c^2_{\vis}$
and $n_s$ worse, thus weakening our constraints. However, a
measurable running is not expected in the most common inflationary
models and there is no strong indication for it from the data we
considered, so it conservative to exclude this possibility from
our analysis. Another possibility would be to consider a dark
energy model different from a cosmological constant. The impact of
dark energy is mainly in the position of the acoustic peaks and on
the large scale CMB spectrum, which is rather orthogonal to the
power suppression signature due to the viscosity parameter. Hence
we do not expect a significant degeneracy between such models and
$c^2_{\vis}$. Finally, we made use of a very limited number of
data-sets. Inclusion of Lyman-alpha data from SDSS, for example,
would probably improve our constraints on the spectral index $n_s$
and therefore provide a tighter constrain on $c^2_{\vis}$. The
latest Ly-$\alpha$ analysis from SDSS \cite{Se04} finds
$n_s=0.98\pm0.02$ which would further close our likelihood
contours around the $c_s^2=1/3$ region. However, the precise
statistical errors to be attached to such measurements are still
under debate among specialists, so we preferred not to include
this data at the moment.

One should note that the method we adopted provides the only way
to detect primordial anisotropies in the neutrino background in
the foreseeable future. Furthermore, this work illustrates the
power of current cosmological data-sets to test the hot Big Bang
model and constrain subtle details of the energy density
components of our universe. For instance, one might speculate that
detecting $c_{\vis}^2\neq1/3$ for the NB would hint to interaction
of neutrinos with other species (one such example has been
recently introduced by \cite{scott} and investigated in
\cite{hannestad}), including exotic components like dark energy.
Future measurements and data releases should be able to strengthen
our result, possibly shedding new light on the physics of the
early universe.

\textit{Acknowledgements} It is a pleasure to thank Scott Dodelson
for comments and Rachel Bean for discussions and help. R.T. is
supported by the Tomalla Foundation and by the Royal Astronomical
Society. A.M. is supported by MURST through COFIN contract no.\
2004027755. The use of the Myrinet cluster owned and operated by
the University of Geneva is acknowledged. We acknowledge the use
of the Legacy Archive for Microwave Background Data Analysis
(LAMBDA). Support for LAMBDA is provided by the NASA Office of
Space Science.


\begin{thebibliography}{99}
\bibitem{Bennett03} C.L. Bennett \etal, \ApJSS {\bf 148}, 1 (2003).
\bibitem{Tg04}  M.~Tegmark {\it et al.} (SDSS Collaboration),
                Phys.\ Rev.\ D {\bf 69}, 103501 (2004).
\bibitem{Se04}  U.~Seljak {\it et al.},
                Phys.\ Rev.\ D {\bf 71}, 103515 (2005).
\bibitem{spergel} D.~N.~Spergel {\it et al.}, \ApJSS {\bf 148}, 175 (2003).
\bibitem{hagmann}
C.~Hagmann,
astro-ph/9905258.
%
\bibitem{Bashinsky:2003tk}
S.~Bashinsky and U.~Seljak,
Phys.\ Rev.\ D {\bf 69}, 083002 (2004).
\bibitem{bowen} R.~Bowen, S.~H.~Hansen, A.~Melchiorri, J.~Silk and R.~Trotta,
Mon.\ Not.\ Roy.\ Astron.\ Soc.\  {\bf 334}, 760 (2002);
S.~Hannestad,
J.\ Cosmol.\ Astropart.\ Phys.\ {\bf 0305}, 004 (2003);
E.~Pierpaoli,
Mon.\ Not.\ Roy.\ Astron.\ Soc.\  {\bf 342}, L63 (2003);
P.~Crotty, J.~Lesgourgues and S.~Pastor,
Phys.\ Rev.\ D {\bf 67}, 123005 (2003).
\bibitem{fogli}
G.~L.~Fogli, E.~Lisi, A.~Marrone, A.~Melchiorri, A.~Palazzo,
P.~Serra and J.~Silk,
Phys.\ Rev.\ D {\bf 70} 113003 (2004).
\bibitem{Huetal95}
W.~Hu, D.~Scott, N.~Sugiyama and M.~J.~White,
Phys.\ Rev.\ D {\bf 52}, 5498 (1995).
\bibitem{Hu98}
W.~Hu,
Astrophys.\ J.\  {\bf 506}, 485 (1998).
\bibitem{Huetal98}
W.~Hu, D.~J.~Eisenstein, M.~Tegmark and M.~J.~White,
Phys.\ Rev.\ D {\bf 59}, 023512 (1999).
\bibitem{Lewis:2002ah}
A.~Lewis and S.~Bridle,
Phys.\ Rev.\ D {\bf 66}, 103511 (2002).
\bibitem{wmap1}
C.L. Bennett \etal, \ApJSS {\bf 148}, 1 (2003);
G. Hinshaw {\it et al.}, \ibid{148}{2003}{135}.
\bibitem{verde} L. Verde {\it et al.}, \ApJSS {\bf 148}, 195 (2003).
\bibitem{cbi}
A.C.S.~Readhead {\it et al.},
Astrophys.\ J.\  {\bf 609}, 498 (2004).
\bibitem{vsa}
C.~Dickinson {\it et al.},
astro-ph/0402498.
\bibitem{acbar1}
J.H. Goldstein {\it et al.}, \apj {\bf 599}, 773 (2003);
C.-l. Kuo {\it et al.}, \ibid{600}{2004}{32}.
\bibitem{acbarpage} Data available from
\texttt{http://cosmologist.info/ACBAR}
\bibitem{Lazarides:2004we}
G.~Lazarides, R.R.~de Austri and R.~Trotta,
Phys.\ Rev.\ D {\bf 70}, 123527 (2004).
\bibitem{sloan}
M.~Tegmark {\it et al.} (SDSS Collaboration),
Astrophys.\ J.\  {\bf 606}, 702 (2004).
\bibitem{freedman}
W.~L.~Freedman {\it et al.},
Astrophys.\ J.\  {\bf 553}, 47 (2001).
\bibitem{bbn}
E.~Lisi, S.~Sarkar and F.~L.~Villante,
  Phys.\ Rev.\ D {\bf 59}, 123520 (1999);
 J.~P.~Kneller, R.~J.~Scherrer, G.~Steigman and T.~P.~Walker,
  Phys.\ Rev.\ D {\bf 64}, 123506 (2001);
R.~H.~Cyburt, B.~D.~Fields and K.~A.~Olive,
  Phys.\ Lett.\ B {\bf 567}, 227 (2003);
A.~Cuoco, F.~Iocco, G.~Mangano, G.~Miele, O.~Pisanti and
P.~D.~Serpico,
  Int.\ J.\ Mod.\ Phys.\ A {\bf 19}, 4431 (2004).
\bibitem{robbays} R.~Trotta,
  astro-ph/0504022.
\bibitem{scott}
  J.~F.~Beacom, N.~F.~Bell and S.~Dodelson,
  Phys.\ Rev.\ Lett.\  {\bf 93}, 121302 (2004).
\bibitem{hannestad}
S.~Hannestad,
JCAP {\bf 02}, 011 (2005).

\end{thebibliography}
\end{document}